\documentclass[aip,rsi,reprint]{revtex4-1}

\usepackage{graphicx, amsmath, amssymb}
\usepackage{dcolumn} 
\usepackage{hyperref}

\begin{document}

\title{Note: Investigation of a Marx generator imitating a Tesla transformer}
\author{B. H. McGuyer}%
\altaffiliation{Present address: Facebook, Inc., 1 Hacker Way, Menlo Park, CA 94025, USA}
\affiliation{Department of Physics, Columbia University, 538 West 120th Street, New York, NY 10027-5255, USA}
\date{\today}%

\begin{abstract}
A compact Marx generator was built to mimic a spark-gap Tesla transformer. 
The generator produced radio-frequency pulses of up to $\pm$200 kV and $\pm$15 A with a frequency between 110 to 280 kHz at a repetition rate of 120 Hz. 
The generator tolerated larger circuit-parameter perturbations than is expected for conventional Tesla transformers. 
Possible applications include research on the control and laser guiding of spark discharges. 
\end{abstract}

\maketitle

Tesla transformers (or Tesla coils) are pulsed-power supplies that generate bursts of radio-frequency alternating current at very high voltages.\cite{naidu:1995}  
They are relatively simple, compact, and inexpensive so have been used in a wide range of applications from particle acceleration to insulation testing. 
Recently, there has been a renewed interest in the ability of Tesla transformers to produce long electrical discharges in air 
because these discharges can be guided by laser filaments.\cite{OptExp:2012:TCETAL,APL:2012:TCETAL,APL:2013:TCETAL,JAP:2014:TCETAL,SciRep:2017:GuidedTCDischargeETAL} 
To date, laser-guided discharges from Tesla transformers have achieved a greater enhancement in length than those from other, primarily direct-current supplies,\cite{houard:2014:reviewETAL}  
and additionally can be produced using only a single electrical terminal.\cite{APL:2012:TCETAL}  
This makes Tesla transformers and similar supplies attractive for research towards the control of electrical discharges,\cite{SciAdv:2015:neatSparkGuidingAroundObjectsETAL} 
their interaction with laser filaments,\cite{APL:2016:prolongguideddischargeETAL} 
the generation of plasma antennas,\cite{APL:2012:filamentedRFantennaETAL}  
and the laser guiding of lightning.\cite{AIP:2012:ForestierETAL, APL:2017:multi-pulseETAL} 

Conventionally, Marx generators are used for these applications. 
Marx generators are pulsed-power supplies that normally generate short pulses of high-voltage direct current instead of radio-frequency current.\cite{naidu:1995} 
Compared to Tesla transformers, Marx generators are more straightforward to engineer and their output is more reproducible. 
Furthermore, Marx generators do not rely on resonant coupling like Tesla transformers, which is sensitive to changes in circuit parameters. 
This sensitivity is a potential limitation for Tesla transformers in these applications because dynamic changes to their load, such as the evolution of spark discharge, can disrupt this resonant coupling during operation.  
Marx generators, in contrast, are nearly immune to this sensitivity by design. 

Ideally, the best aspects of these two power supplies could be combined in an improved supply. 
This Note demonstrates that Marx generators can be designed to imitate Tesla transformers, combining an output similar to that of a Tesla transformer with the circuit architecture of a Marx generator. 
Such modified Marx generators are shown to tolerate circuit-parameter changes more than Tesla transformers, making them attractive alternatives to Tesla transformers in the applications mentioned above. 
While compact,\cite{arantchoukMarx:2014ETAL} high-repetition,\cite{IEEE:2010:100HzMarxETAL} and inductively loaded\cite{APL:2016:prolongguideddischargeETAL} Marx generators and coupled Marx-Tesla circuits\cite{APL:1978:MarxTesla} exist, this work demonstrates that a modified Marx generator can mimic a conventional, loosely coupled spark-gap Tesla transformer (SGTT).

\begin{figure}[b] 
	\centering
	\includegraphics[width=8.5cm]{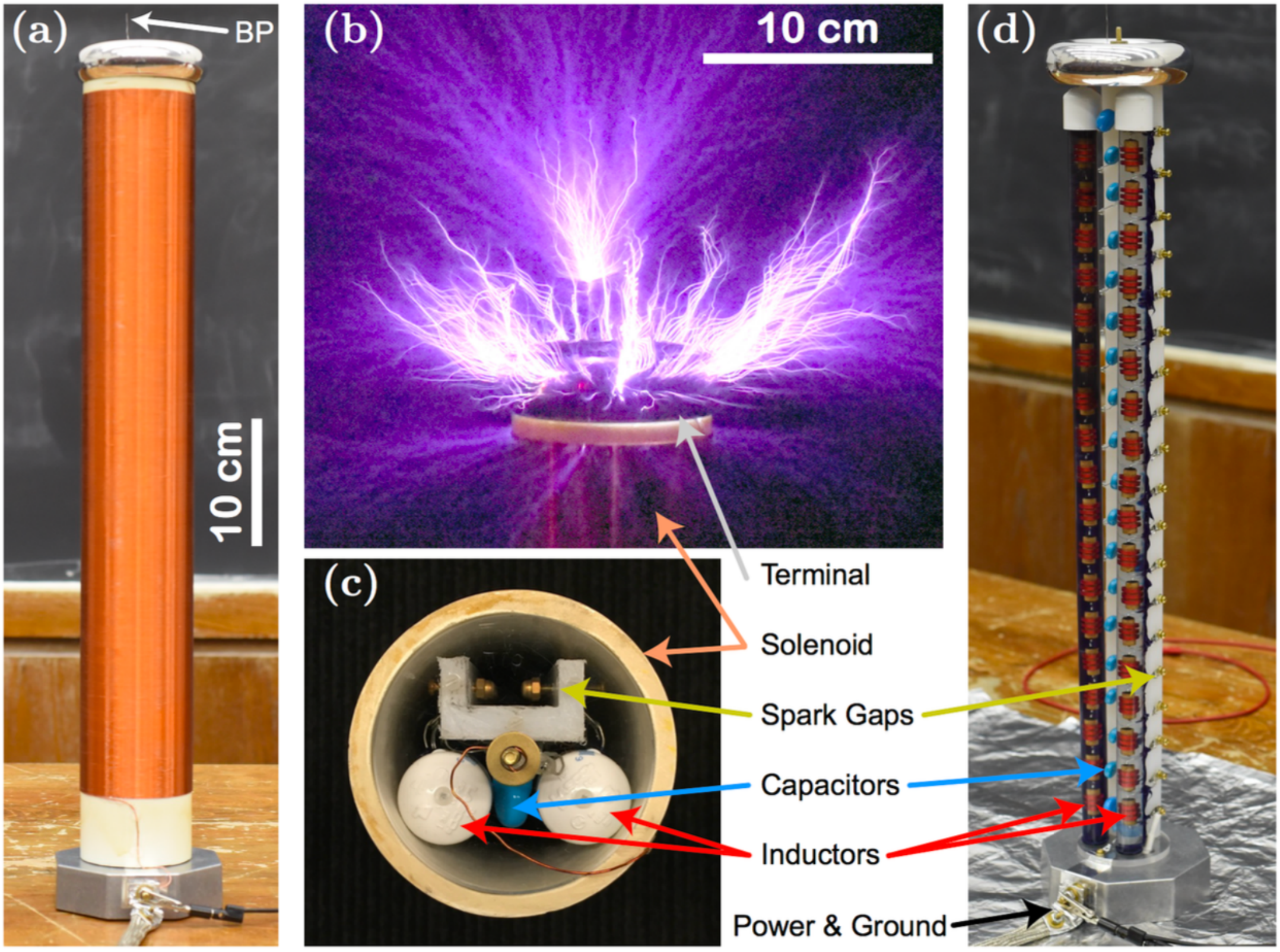}
	\caption{
	Marx generator imitating a spark-gap Tesla transformer. 
	(a) Side view showing a metal terminal above a single-layer solenoid inductor and a metal base with power and ground wiring. 
	BP denotes a breakout point to aid spark emission. 
	(b) Spark discharge emitted from the terminal (1/4 s exposure). 
	(c) View beneath the terminal showing the Marx generator inside the solenoid. 
	(d) Side view without the solenoid, showing the Marx generator above a ground plane.} 
	\label{fig1}
\end{figure}

Fig.~1 shows the modified Marx-generator apparatus, hereafter a Marx coil (MC), which resembles a compact Tesla transformer without a primary coil (omitted because there is no resonant coupling). 
During operation, it produces repetitive pulses of high voltage at radio frequencies. 
Like a SGTT, it can be adjusted to produce no, few, or multiple single-ended spark discharges in air depending on the output terminal and power supply configuration. Additionally, it is able to repeatedly breakdown discharge channels from previous pulses, just like SGTTs, as shown by the subtle ``banjo'' effect of comb-like discharges in Fig.~1(b) and in additional photos and video in the supplementary material. 

\begin{figure}[t] 
	\centering
	\includegraphics[width=8.5cm]{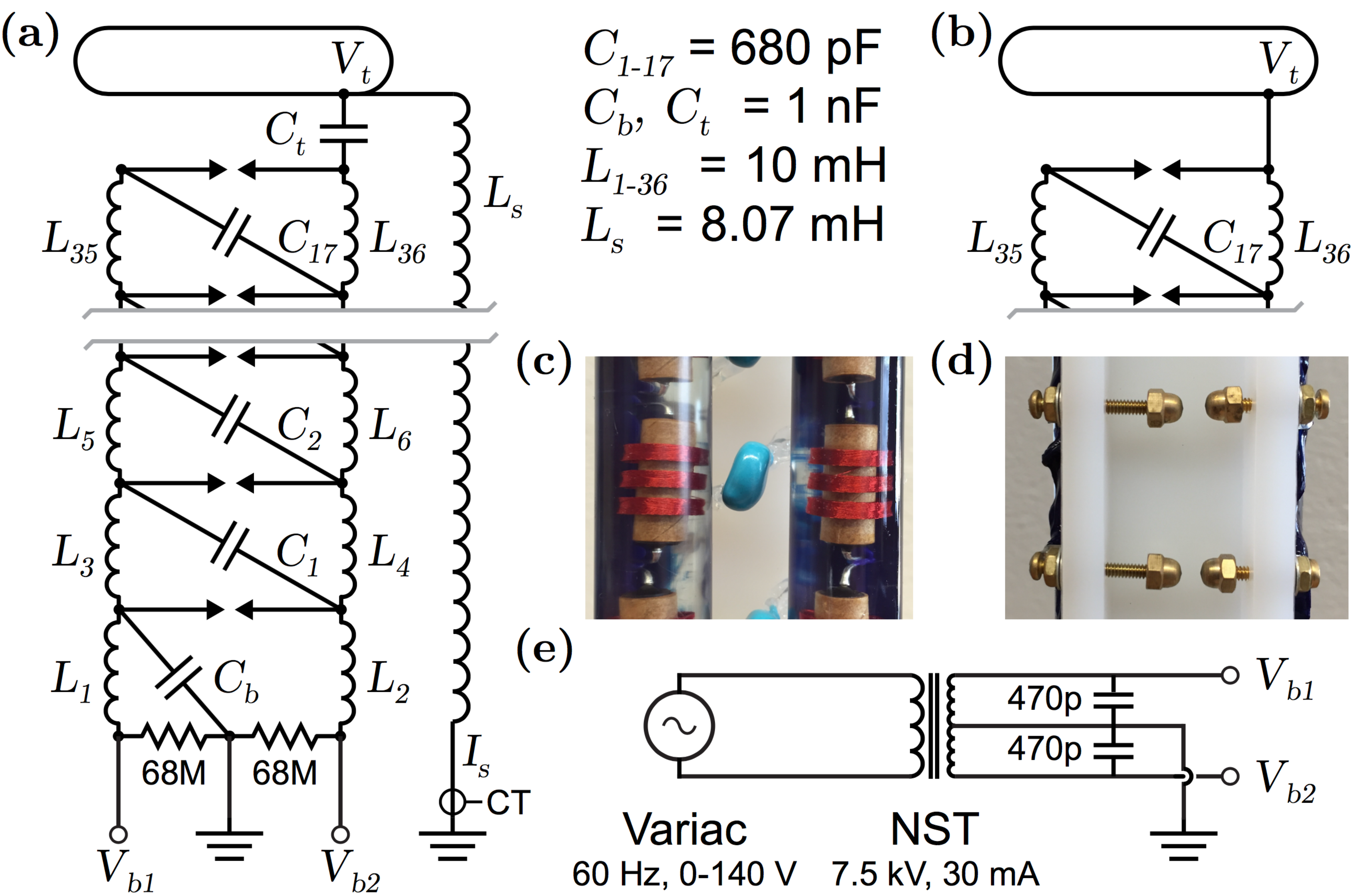}
	\caption{
	Circuit details of the Marx coil apparatus. 
	(a) Component wiring for operation with the solenoid, as in Fig.~1(a). 
	CT denotes a current transformer. 
	(b) Modification to operate without the solenoid, as in Fig.~1(d). 
	(c) Inductors (red) and capacitor (blue) for one Marx stage. 
	(d) Spark gaps. 
	(e) Power supply. 
	The variac was set to 140 V$_\text{rms}$ for all data shown. 
	Additional  details are in the supplementary material. 
	} 
	\label{fig2}
\end{figure}

\begin{figure}[t] 
	\centering
	\includegraphics[width=8.5cm]{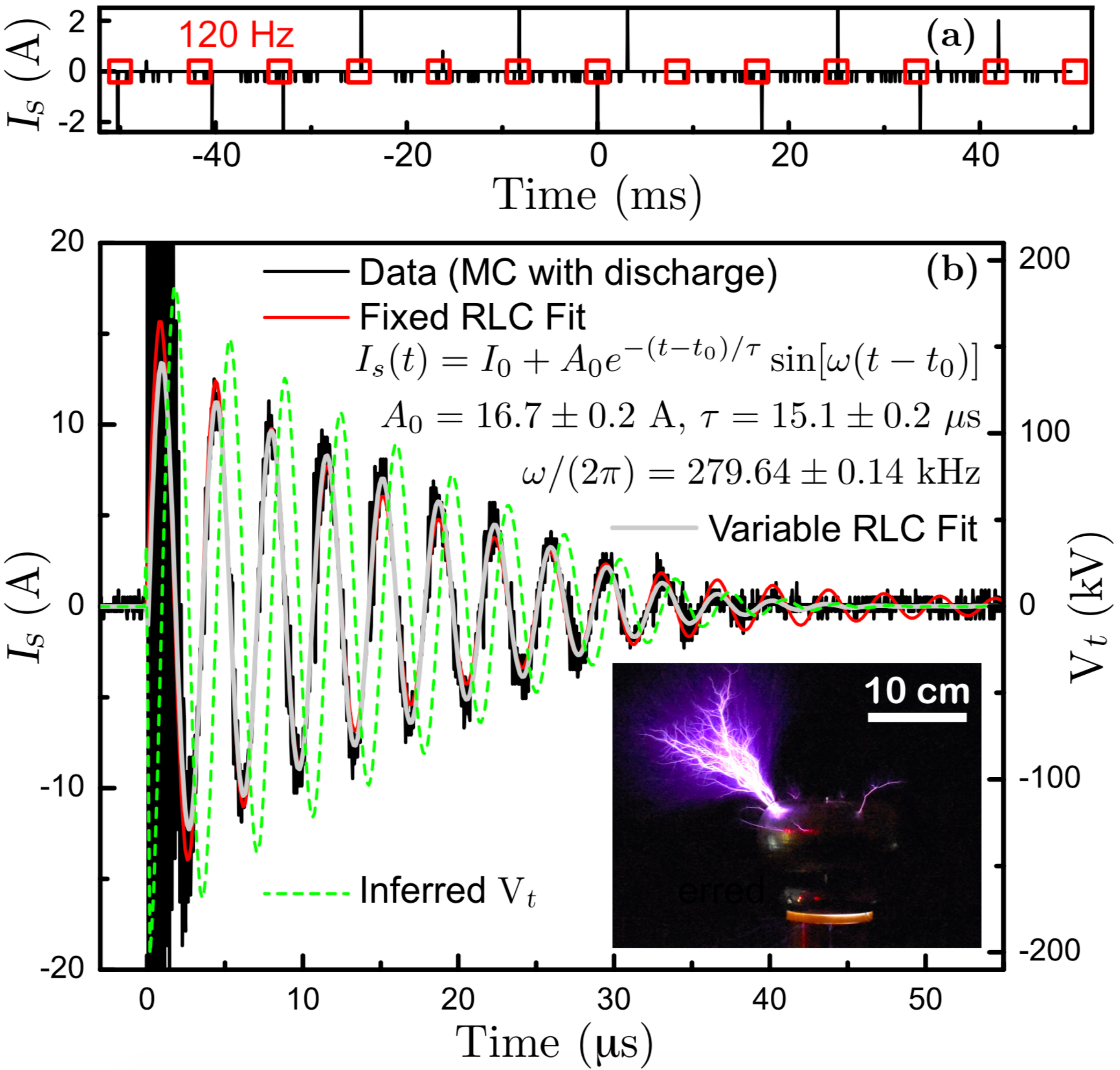}
	\caption{ 
	Output during operation with the solenoid. 
	(a) Like a SGTT, the Marx coil produced pulses at a roughly 120 Hz rate, which appear as under-sampled 
	lines in the solenoid current $I_s$ measured by a current transformer. 
	(b) Current $I_s$ and inferred voltage $V_t \approx - L_s \, d I_s/dt$ during a pulse. 
	The decaying exponential oscillation of an RLC circuit fits $I_s$ well, but allowing time-varying RLC parameters improves the fit (see supplementary material). 
	Inset shows representative spark discharge for this data, which used a larger terminal and a different breakout point than shown in Fig.~1(a). 
	} 
	\label{fig3}
\end{figure}

While the secondary solenoid of a classic SGTT has no components inside, here the solenoid contains a Marx generator as shown in Fig.~1. 
A plastic U-channel provides a backbone for the Marx generator components that form the circuit in Fig.~2. 
Inductors are used instead of resistors for fast charging that is enhanced by the solenoid. 
Additionally, the solenoid provides electric-field grading to reduce stresses on the components inside and suppresses light and sound emission. 
To prevent flashover, the inductors are immersed in mineral oil and the capacitor leads insulated with silicone. 
The MC is charged by alternating current (AC) from a neon sign transformer (NST). 
NSTs are convenient here and widely used for SGTTs because their current-limited output tolerates short circuiting. 

The spark gaps are adjusted so that when the charging voltage is near a maximum the gaps close to erect the Marx generator. 
Just as in a SGTT, this leads to a pulsed output that repeats at roughly twice the NST AC frequency, or 120 Hz, made of bursts of radio-frequency high voltage. 
Fig.~3 shows a typical waveform captured by measuring the solenoid base current $I_s$. 
The burst in Fig.~3(b) is similar to that of a SGTT powered by the same NST, though not identical. 
Here, the waveform is approximated well by the exponentially decaying oscillation of an RLC circuit and does not have the slowly modulated (``beating'') envelope typical of a SGTT. 
The red curve is a fit assuming fixed RLC parameters. 
The solenoid current is nearly spatially uniform, unlike in some SGTTs, 
allowing the top voltage to be estimated from the base current as $V_t \approx - L_s \, d I_s/d t$. 
The inferred peak voltage $|V_t|$ was 201 kV using the variable RLC fit. 

In principle, a Marx generator can be adjusted to mimic a SGTT as follows. 
During operation, a Marx generator charges $N$ stages each with capacitance $C_0$ in parallel and rapidly rewires the stages in series to produce a pulse. Ideally, the maximum output voltage is $N$ times the charging voltage. 
In contrast, a SGTT charges the capacitance $C_p$ of a primary oscillator circuit and then transfers this energy via resonant coupling to a secondary oscillator circuit with capacitance $C_s$. 
From energy conservation, the maximum possible output voltage is $\sqrt{C_p/C_s}$ times the charging voltage.\cite{naidu:1995} 
Therefore, for the same charging voltage, choosing $C_0 = C_p/N$ leads to the same energy per pulse, and choosing $N = \sqrt{C_p/C_s}$ 
to roughly the same output voltage. 
To produce an oscillatory output like a SGTT, the Marx generator then needs a suitable inductance in parallel with the total erected capacitance, which may come from either an inductive load, the stage impedances, or both. 
Choosing the same output frequency leads to a similar output impedance, depending on the spark gap and component losses. 

The MC in Fig.~1 was designed by first selecting a power supply and repetition rate common for a compact SGTT. 
Then the charging capacitance was chosen to be near the maximum set by the power supply and rate, which limit the energy per burst. 
The number of stages $N = 18$ was chosen to be less than that ($\sim$ 35) matching a comparable SGTT to enable the voltage measurement in Fig.~3. 
($C_b$ and $C_t$ act as the 18th stage.) 
The solenoid inductance was then chosen to produce an output oscillation frequency typical of SGTTs. 
The inductances of the lossy stage inductors were chosen through SPICE simulation to 
optimize charging speed versus pulse duration.

The sensitivity of Tesla transformers to changes in circuit parameters comes from the resonant coupling that transfers energy between their primary and secondary circuits.\cite{naidu:1995,skeldon:1997ETAL} 
Changes that shift the resonant frequency $f_0$ of either circuit away from their intended values will degrade performance unless the shift $\delta f_0$ is roughly within the coupling bandwidth, or approximately $f_0/Q$ using the quality factor Q of the lossy primary. This leads to the rough limit $|\delta f_0 / f_0| \lesssim 1/Q$, beyond which the shift impedes energy transfer. 
Quantitatively, the curve in Fig.~3(b) corresponds to $Q \approx \omega \tau / 2 = 13.3 \pm 0.2$, which is similar to a typical SGTT value.\cite{skeldon:1997ETAL}  
Thus for a comparable SGTT, the limit 
$|\delta f_0 / f_0| \lesssim$ 7.5\%. 

As a result, slightly adjusting the capacitance or inductance of the secondary, for example, typically ruins SGTT performance and either reversing or compensating for this in the primary is needed to restore operation. 
In contrast, both may be adjusted freely without requiring any other circuit changes to maintain operation with the Marx coil. 
As a demonstration, Fig.~4 shows the MC operating after the solenoid was removed, as in Figs.~1(d) and 2(b). 
This reduced the output frequency to about 110 kHz, corresponding to $\delta f/f_0 \approx -61$\%, far outside the rough limit of $\pm$7.5\%.  
Additionally, this led to spark discharge off the MC components 
due to the lack of field grading by the solenoid, as shown in Fig.~4.

\begin{figure}[t] 
	\centering
	\includegraphics[width=8.5cm]{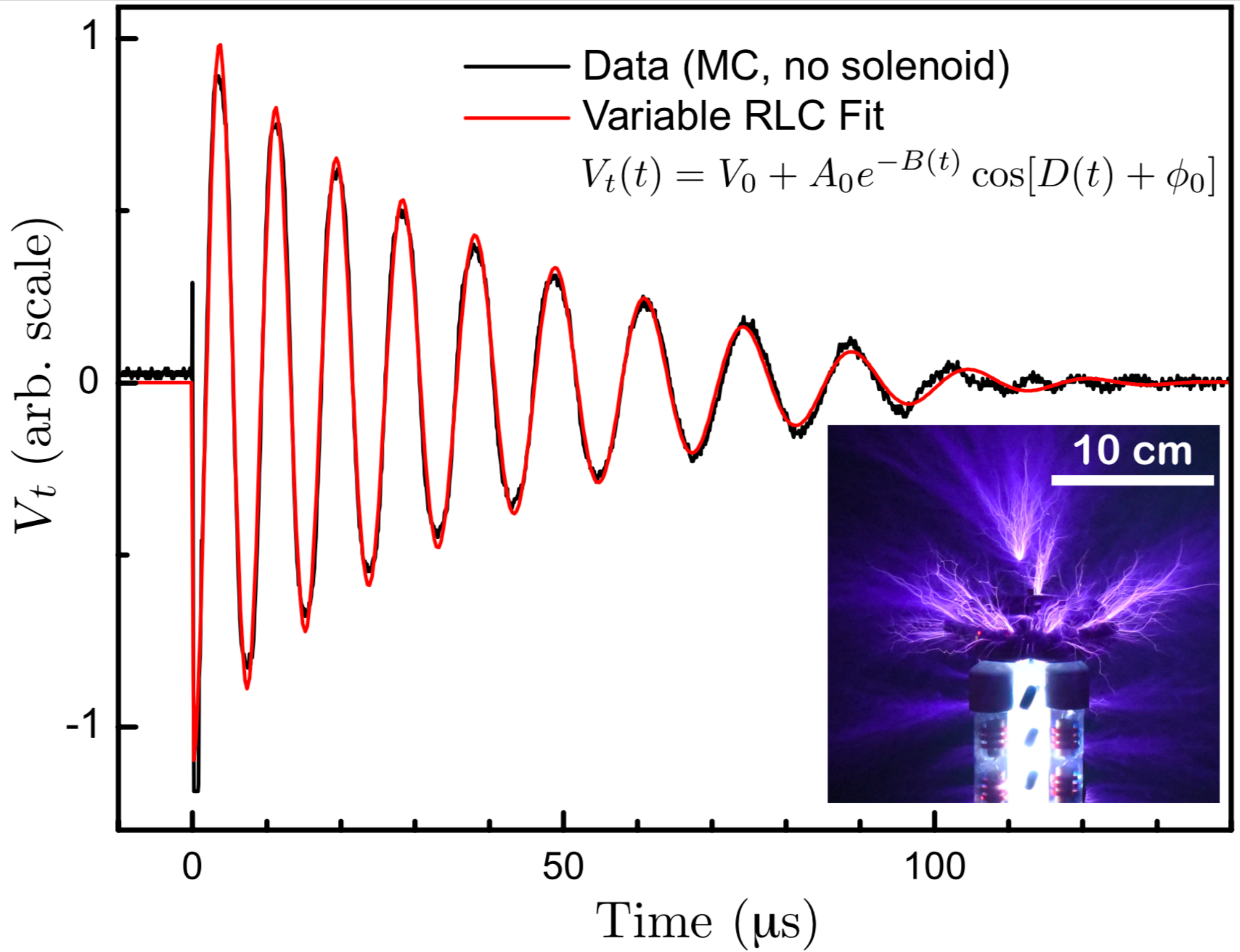} 
	\caption{
	Output during operation without the solenoid. 
	The voltage $V_t$ oscillates with a frequency that ramps from roughly 140 to 95 kHz, as measured by an uncalibrated capacitive-pickup probe. 
	Here, the data require time-varying RCL parameters to be fit well (see supplementary material), in contrast to Fig.~3(b) that has a more subtle ramp. 
	Inset shows representative spark discharge for this data.  Without the solenoid, discharge also occurs off the components. 
	} 
	\label{fig4}
\end{figure}

In addition to such static changes, SGTT circuit parameters can also change dynamically. 
In this case, the rough limit given above holds approximately, although it ignores  
possibly beneficial effects like rapid adiabatic passage.\cite{RAP} 
In contrast, the data in Fig.~4 show a clear frequency ramp from about 140 to 95 kHz, 
corresponding to $\delta f_0/f_0 \approx -32$\%, 
highlighting that the MC tolerates dynamic changes. 
Similar frequency ramps were observed in all data including that in Fig.~3(b), for which it is more subtle. 
Field-sensitive ceramic stage capacitors are likely responsible for the ramp in Fig.~3(b), and together with stage-inductor saturation are likely responsible for the ramp in Fig.~4.

While the sources of dynamic changes observed here can be removed by replacing components, other sources may be unavoidable. 
In particular, the development and evolution of transient spark discharge dynamically loads supplies like SGTTs. 
For example, growing a long leader-like structure 
effectively loads the supply with $\sim$ 3 pF per meter of length.\cite{raizer}  
Unfortunately, no reproducible trend was observed 
that could be attributed to spark discharge, likely because of a larger variability in component effects. 

This electrical loading from spark discharge is one potential obstacle to future research with Tesla transformers towards the laser guiding  of long sparks, because 
their output capacitance is typically small ($\sim$20--50 pF). 
Unfortunately, the effects of such discharge loading on Tesla transformers have not been extensively studied (see supplementary material). 

In summary, a Marx generator was modified to mimic a Tesla transformer, producing similar output and spark discharge. This apparatus tolerated larger changes in its circuit parameters than is expected for Tesla transformers. Thus, such supplies may be attractive alternatives to Tesla transformers in research with the production, control, and laser guiding of spark discharges.

\section*{Supplementary Material}
See supplementary material for additional apparatus, analysis, and spark discharge details (including video). 

\begin{acknowledgments}
I am grateful to 
Tanya Zelevinsky and Ken Sikes for their support with equipment and space, 
to Taylor Chapman for photography, 
and to Mickey McDonald, Ben Olsen, and Steve Ward for reading the manuscript. 
This work was performed before the author joined Facebook. 
\end{acknowledgments}

\begin{figure}[t!] 
	\centering
	\includegraphics[width=8.6cm]{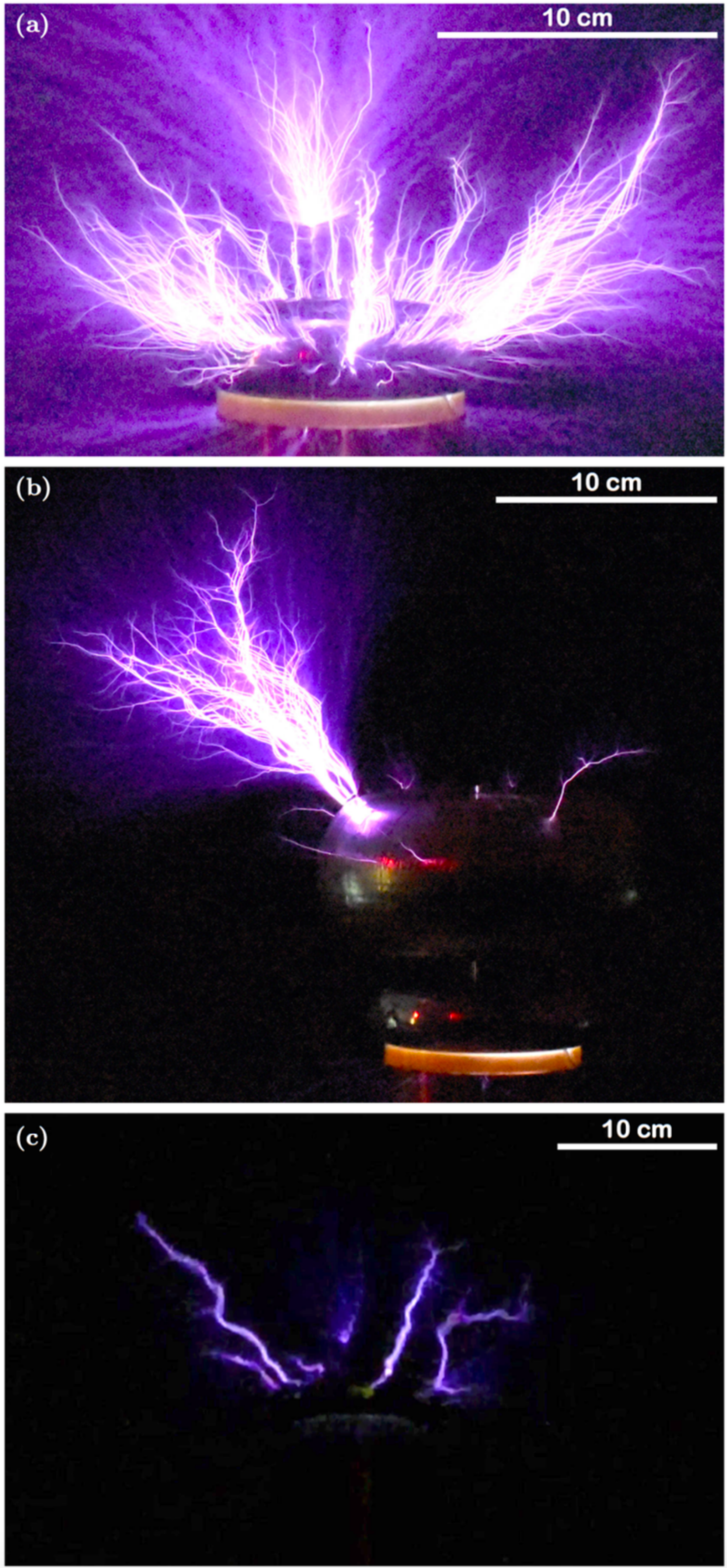}
	\caption{
	Spark discharge photographs and video. 
	(a) Larger version of inset in Fig.~1.  
	A vertical, thin metal wire acted as a breakout point for several sparks that do not appear to connect to the output terminal. 
	(b) Larger version of inset in Fig.~3, with a larger terminal placed above the terminal shown in (a). 
	A piece of metal foil tape acted as a breakout point for the longest discharges. 
	(c) Still image from supplementary video. 
	The terminal arrangement is the same as in (a).} 
	\label{figSI-banjo}
\end{figure}

\section{Supplementary Material}

\subsection{Spark discharges}

Fig.~5 provides larger versions of the insets showing spark discharge in Figs.~1 and 3. 
These pictures and the supplemental video 
show that the Marx coil apparatus is able to re-breakdown previous discharge channels from past bursts, as is known to occur with Tesla transformers. 
This phenomenon is visible in photographs and videos because of convection during the time between bursts (about 8.3 ms here), similar to how convection produces a rising spark structure in a Jacob's ladder. 
In the photographs, it leads to comb-like discharge patterns, or the so-called ``banjo effect,'' because bursts briefly illuminate each rising channel during the exposure. 
This process is similar to streak photography except that here the spark discharge is moving instead of the camera system.  
In the video, this is shown by the illusion of vertically rising discharge structures that evolve over time. 

During testing of the Marx coil (MC) apparatus, the longest observed discharge lengths were roughly 15 cm 
for free discharge in air for operation with the solenoid as arranged in the inset of Fig.~3, 
and 20 cm  
for discharges from a wire to a flat grounded target without the solenoid as arranged in Fig.~4.

\subsection{Marx-coil aparatus}

The total height of the MC apparatus as shown in Fig.~1(a) was roughly 68 cm. 
The capacitors $C_{1-17}$, $C_b$, and $C_t$ were Murata DHR series ceramic disc capacitors with a ZM temperature characteristic, $\pm 10$\% tolerance, and 15 kV direct-current (DC) rating. 
Their capacitance is known to decrease with DC bias, up to roughly 22\% at 15 kV.\cite{murata} 

The inductors $L_{1-36}$ were 3-pi universal wound Bourns/J.~W.~Miller 6306-RC varnished RF chokes with ferrite cores, $\pm 5$\% tolerance, and 31 $\Omega$ or less DC resistance. 
These inductors were installed in one of two clear plastic tubes, each with a slot for the leads that was later sealed with room-temperature-vulcanizing (RTV) silicone, and the inductors were immersed in mineral oil. 

The load inductor $L_\text{s}$ was a close-wound single-layer solenoid made from a 55.4 cm varnished winding of approximately 791 turns of 22 awg magnet wire on a plastic pipe with an outer diameter of 8.8 cm. The measured inductance was 8.07 $\pm$ 0.03 mH at 10 kHz, and DC resistance was 11.8 $\pm$ 0.1 $\Omega$. 
The estimated effective Medhurst self-capacitance\cite{lee} is $\sim$ 8 pF. 

Note that coupling is expected between the load and stage inductors as installed. This coupling could be suppressed with different physical arrangements, or by using different winding patterns.\cite{FieldContainingInductors} Alternatively, the load and stage inductors need not be separate components. 

The solenoid base current was measured using a Pulse F15155NL current transformer with a 1 k$\Omega$ shunt resistor, which gave a sensitivity of  2 V/A over 0.5--700 kHz. 
The bandwidth limitations of the current transformer led to the initial measurement error in Fig.~3(b), and of a capacitive-pickup probe to the same (though less visible) in Fig.~4. 

The spark gap electrodes were made of brass 4-40 acorn nuts with a radius of curvature of about 2.5 mm. 
The gap spacings used were roughly 
3.0, 3.0, 3.0, 3.0, 3.0, 3.0, 3.5, 4.0, 4.5, 5.0, 5.5, 6.0, 6.5, 7.0, 7.5, 8.0, 8.0, and 8.0 mm, from bottom to top. 

When installed, the plastic pipe form of the solenoid muffled and blocked the spark gap noise and light emission significantly. White plastic was used for the U-channel and solenoid pipe so that the interior space resembled an integrating cavity, in case it might  reduce gap jitter and losses. 
Note that gap quenching is not critical here unlike in spark-gap Tesla transformers (SGTTs), since energy is not transferred between resonant circuits. 

Two different toroidal output terminals were used for the data shown. 
The first, shown in Figs.~1 \& 4, had a maximum width of 
4.8 cm, 
height of 2.9 cm, 
and an estimated electrostatic capacitance of 5.5 pF. 
In Fig.~3, a second terminal was placed above the first terminal. This terminal had a maximum width of 15.3 cm, height of 7.0 cm, and an estimated electrostatic capacitance of 6.0 pF. 
A vertical metal wire segment was used as a breakout point for spark production in Figs.~1 and 4, and a folded piece of metal foil in Fig.~3. 

A maximum value $C_\text{max}$ for the charging capacitance can be roughly estimated from the power supply parameters and the repetition rate of 120 Hz as follows. Consider a 60 Hz AC power supply with rated output power $P = V_\text{rms} I_\text{rms}$ and output impedance $Z = V_\text{rms}/I_\text{rms}$, where $V_\text{rms}$ and $I_\text{rms}$ are the maximum root-mean-square output voltage and current. Then $C_\text{max}$ corresponds to the capacitance that charges from zero voltage up to $\sqrt{2}\, V_\text{rms}$ during a quarter cycle, or 1/240 second, at the rated power. From energy conservation this requires $C_\text{max} V_\text{rms}^2 = P / (240~\text{Hz}),$ which gives 
\begin{align}	\label{LTR}  
C_\text{max} = 1/(Z \times 240~\text{Hz}). 
\end{align}  
For the NST of Fig.~2(e), $C_\text{max} \approx 16.7$ nF. 
As built, the effective charging capacitance for the arrangement of Fig.~2(a) was 12.1 nF, 
and of Fig.~2(b) was about 11.8 nF, ignoring the NST filter capacitors (which added $\sim$ 0.24 nF).  

In practice, the optimum charging capacitance that leads to the most energy per burst without significantly reducing the charging voltage should be found empirically, because NSTs are not ideal AC power supplies. 
Note that in practice charging may occasionally occur for longer (or shorter) than a quarter cycle, depending on the spark gap and load behavior, which can lead to larger charging voltages than the power supply rating. 

The estimated erected Marx bank capacitance for the arrangement of Fig.~2(a) with the solenoid is 37 pF, and for that of Fig.~2(b) without the solenoid is 38.5 pF. 
In practice, the total erected capacitance will include additional capacitance from the terminal and solenoid, as estimated above (neglecting their interaction), and to the environment. 
Approximating the NST as a short during a burst, the erected Marx inductance for Fig.~2(a) is roughly 7.4 mH  
and for Fig.~2(b) without the solenoid is roughly 90 mH. 

\subsection{Estimating time-dependent circuit parameters}
During operation, the apparatus behaves approximately like a damped, undriven RLC circuit with time-dependent parameters. 
This is because during a burst the spark gaps act approximately as short circuits, making the circuit in Fig.~2 resemble that of Fig.~6. 
The circuit parameters and their time dependence can be estimated from ``variable RLC fits'' to the data in the following manner. 

\begin{figure}[t] 
	\centering
	\includegraphics[]{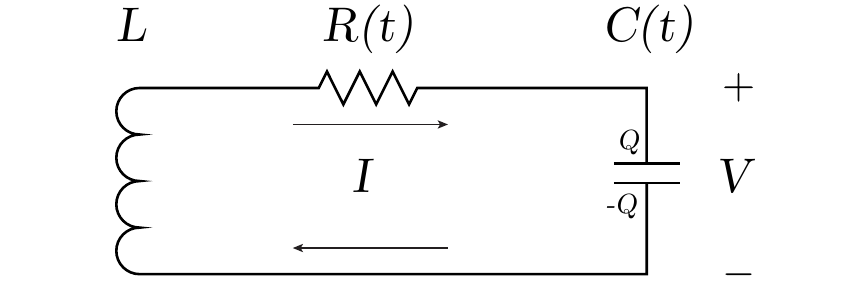}
	\caption{Lumped series RLC circuit with constant inductance $L$ and time-dependent resistance $R(t)$ and capacitance $C(t)$.} 
	\label{fig:varRLC}
\end{figure}

Consider a lumped series RLC circuit as sketched in Fig.~6.  
Let us assume that the inductance $L$ is constant, but that the resistance $R = R(t)$ and capacitance $C = C(t)$ may vary in time. 
This follows from noting that the inductance is set mainly by the inductors $L_1$ to $L_{36}$ and $L_s$, which are not expected to vary during operation except when the solenoid $L_s$ is removed, in which case the increased current through the stage inductors may lead to saturation. 
Following Kirchoff's voltage law, the circuit satisfies the differential equation 
\begin{align}
(LI)'+ R I + {Q}/{C} = 0. 
\end{align}
Here and subsequently a prime denotes differentiation with respect to time $t$.  
Using this with 
$I = Q'$, 
the charge $Q(t)$ then evolves according to the differential equation 
\begin{align}	\label{RLC_Q}
Q'' + \left( \frac{R}{L} \right) Q' + \left( \frac{1}{LC} \right) Q = 0, 
\end{align}
which has the same form as that for an RLC circuit with constant parameters. 
The current and voltage, however, will not evolve according to this equation.  
Instead, multiplying (\ref{RLC_Q}) by $LC$ and differentiating gives the corresponding equation for the current $I(t)$, 
\begin{align}	\label{Ide} 
I'' + \left( \frac{R}{L} + \frac{C'}{C} \right) I' + \left( \frac{1 + R C'  + R' C}{LC} \right) I  = 0. 
\end{align}
Similarly, the voltage $V(t)$ across the capacitance $C(t)$ evolves according to 
\begin{align}	\label{Vde} 
V'' + \left( \frac{R}{L} + \frac{2 C'}{C} \right) V' + \left( \frac{1 + R C' + L C''}{LC} \right) V = 0, 
\end{align}
which follows from (\ref{RLC_Q}) using $Q = CV$ and assuming $C(t) \neq 0$. 

To determine the circuit parameters from measurements of the current $I(t)$ or voltage $V(t)$, the differential equation was approximately reconstructed by fitting the data with the trial function 
\begin{align}	\label{trialfn} 
y(t) = A_0 e^{- B(t)} \cos[D(t) + \phi_0], 
\end{align}
where the fit parameters $A_0$ and $\phi_0$ are constant but the functions $B(t)$ and $D(t)$ depend on time.  
This form assumes the burst start time is known, and that there is no background offset (which was removed by fitting with an offset parameter). 
This trial function satisfies the differential equation 
\begin{align}	\label{recoveredDE}
y'' + U(t) \, y' + W(t) \, y  = 0
\end{align}
where the functions $U(t)$ and $W(t)$ may be computed as 
\begin{align}	\label{Ut}
U(t) &= 2 B' - {D''}/{D'}   \\
W(t) &= \left( B' \right)^2 + \left( D' \right)^2 + B'' - {B' D''}/{D'}. 
\end{align} 
The data presented here was described well by the trial function (\ref{trialfn}) using the polynomials 
\begin{align}	\label{Bt}
B(t) &= \alpha_1 t + \alpha_2 t^2 + \alpha_3 t^3 \\ 
D(t) &= \omega_1 t + \omega_2 t^2 + \omega_3 t^3, \label{Dt} 
\end{align}
where the fit parameters $\alpha_i$ and $\omega_i$ are independent of time. 
In contrast, an RLC circuit with constant parameters is described by the polynomials 
$B(t) = t \, R/(2L)$ and $D(t) = t \sqrt{1/(LC)^{2} - R^2/(2L)^2}$.

After fitting data with the trial function (\ref{trialfn}), the reconstructed differential equation (\ref{recoveredDE}) may be used with the appropriate differential equation from above to estimate the time-dependent circuit parameters $C(t)$ and $R(t)$. 
For the data presented here, this analysis was simplified using the following approximations.  
For the current data, the approximation 
\begin{align}	\label{Iapprox} 
\left| R C' + R' C \right| \ll 1
\end{align}
simplifies the differential equation (\ref{Ide}) to 
\begin{align}
I'' + \left( \frac{R}{L} + \frac{C'}{C} \right) I' + \left( \frac{1}{LC} \right) I  \approx 0. 
\end{align}
Using this with (\ref{recoveredDE}), the circuit parameters may be estimated using $L$, (\ref{Ut}--\ref{Dt}), and the fit parameters $\alpha_i$ and $\omega_i$ as 
\begin{align} \label{CtI} 
C(t) &\approx 1/[{L \, W(t)}] \\
R(t) &\approx L \left[ U(t) + {W'(t)}/{W(t)} \right]  \label{RtI} 
\end{align}
within the fitted range of times while $I(t)$ may be distinguished from noise.  
Likewise, for the voltage data the approximation 
\begin{align}	\label{Vapprox} 
\left| R C' + L C'' \right| \ll 1 
\end{align}
simplifies the differential equation (\ref{Vde}) to 
\begin{align}
V'' + \left( \frac{R}{L} + \frac{2C'}{C} \right) V' + \left( \frac{1}{LC} \right) V \approx 0. 
\end{align}
In this case, the circuit parameters may be estimated as 
\begin{align} \label{CtV}
C(t) &\approx 1/[{L \, W(t)}] \\
R(t) &\approx L \left[ U(t) + {2 W'(t)}/{W(t)} \right], \label{RtV} 
\end{align}
again, within the fitted range of times while $V(t)$ may be distinguished from noise.  
For both cases, after estimating $C(t)$ and $R(t)$ the initial approximation, either (\ref{Iapprox}) or (\ref{Vapprox}), must be tested for consistency.

\begin{table}[t!]
\caption{\label{tab:1}
Fixed RLC fit parameters for data in Figs.~3 and 7 using the form (\ref{FixedRLC}). 
The offset $I_0$ is excluded. Values in parenthesis are uncertainties in the last digits.
The variation between parameters with and without discharge is comparable to that observed between data sets with the same conditions.} 
\begin{ruledtabular}
\begin{tabular}{l | c c c c}
{Data set} & $A_0$	& $\tau$	& $\omega/(2\pi)$ 	& $t_0$ \\ 
	& Amp 	& $\mu$s	& kHz			& $\mu$s \\
\hline
MC with discharge 	& 16.7(2)	& 15.1(2)	& 279.64(14)		& $-$0.030(6) \\ 
MC without discharge	& 14.4(2)	& 13.5(2)	& 277.96(15)		& $-$0.036(6)  
\end{tabular} 
\end{ruledtabular} 
\end{table}

\begin{figure}[t!] 
	\centering
	\includegraphics[width=8.5cm]{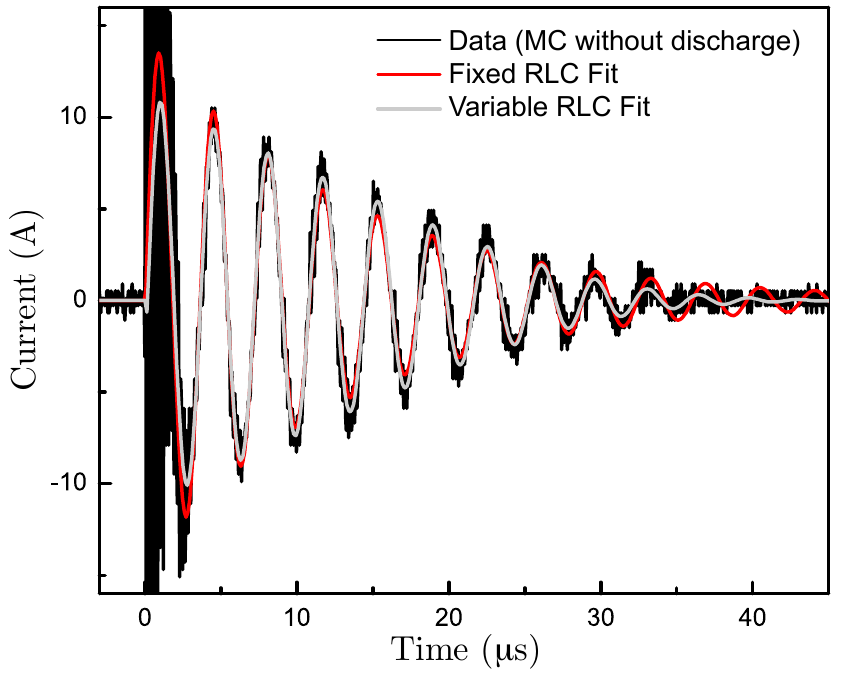}
	\caption{
	Current data and fits corresponding to the ``MC without discharge'' curves of Figs.~9(a) and 10 and parameters of Tables~\ref{tab:1} and \ref{tab:2}.} 
	\label{figSI-Data11}
\end{figure}

\begin{figure}[t!] 
	\centering
	\includegraphics[width=8.5cm]{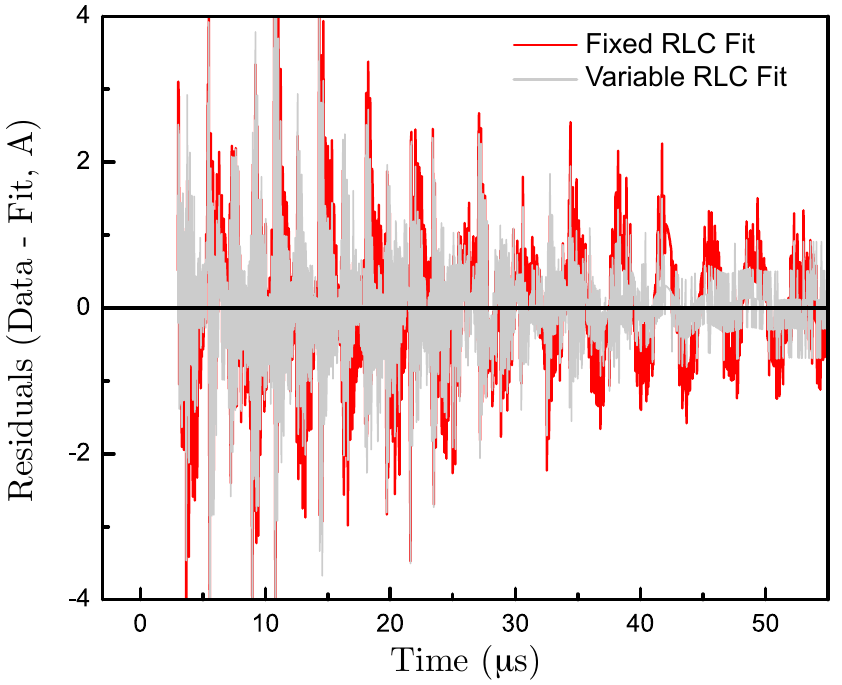}
	\caption{
	Fit residuals for Fig.~3.} 
	\label{figSI-FitResiduals}
\end{figure}

\begin{figure}[t!] 
	\centering
	\includegraphics[]{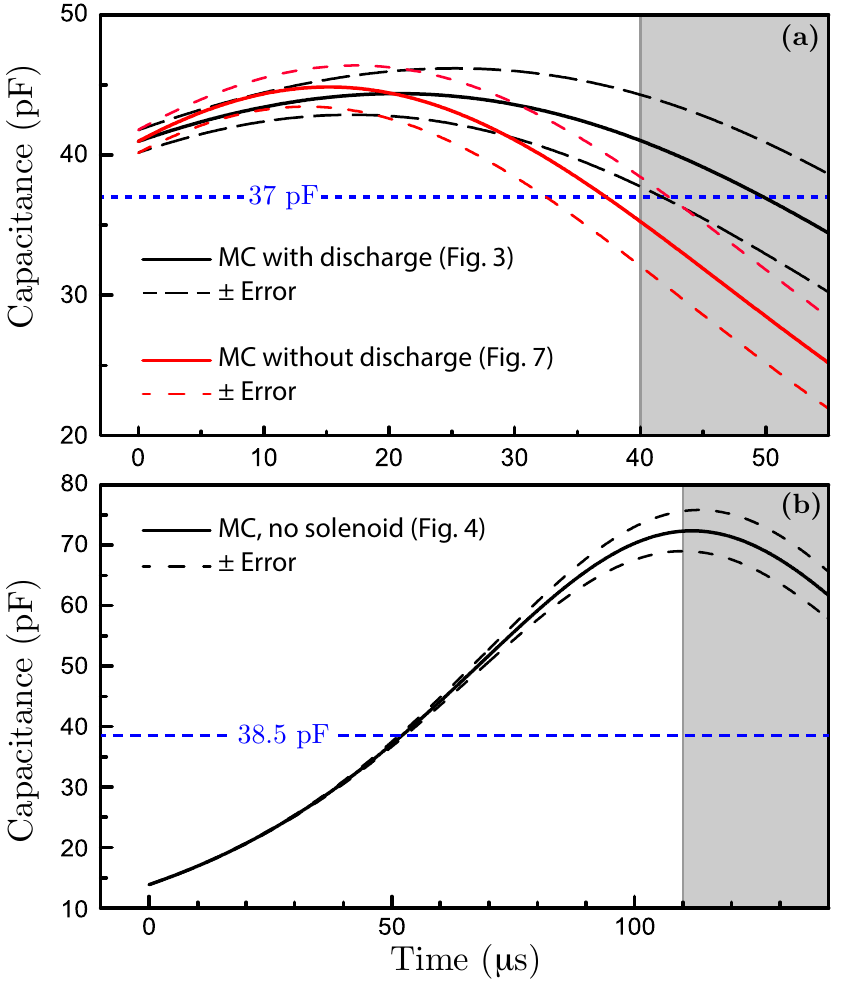}  
	\caption{
	Estimated capacitance from variable RLC fits of measured current or voltage waveforms. 
	(a) MC operation with the solenoid. 
	The black curve is from the data in Fig.~3(b), while the red curve is from data in Fig.~7 in similar conditions but without visible discharge (the breakout point was removed). 
	(b) MC operation without the solenoid. The curve is from the data in Fig.~4 and is likely affected by inductor saturation effects that violate the assumption of constant inductance. 
	Dashed curves show the error from least-squares fit-parameter uncertainties. 
	Horizontal blue dashed lines are the estimated erected Marx bank capacitances, without solenoid and terminal contributions. 
	Grey regions mark when the waveforms are comparable to noise, and the capacitances less reliable. 
	} 
	\label{figSI-OldFig5}
\end{figure}

\begin{table*}[t!]
\caption{\label{tab:2}
Variable RLC fit parameters for data in Figs.~3, 4, and 7 using the form (\ref{VariableRLC}) with expansions (\ref{Bt}) and (\ref{Dt}). 
The offset $y_0$ is excluded. Values in parenthesis are uncertainties in the last digits. 
The variation between parameters with and without discharge is comparable to the variation observed between different data sets with the same conditions.} 
\begin{ruledtabular}
\begin{tabular}{l | c c c c c c c c}
{Data set} & $A_0$	
		& $\alpha_1 \times 10^3$	& $\alpha_2 \times 10^4$	& $\alpha_3 \times 10^6$	
			& $\omega_1$			& $\omega_2  \times 10^3$	 & $\omega_3  \times 10^6$	
				& $\phi_0$	\\ 
	& Amp 
		& MHz	& MHz$^{2}$	& MHz$^{3}$ 	
			& rad MHz 	& rad MHz$^{2}$ 	& rad MHz$^{3}$ 	
				& degree \\
\hline
MC with discharge	& 14.2(5) 
		& 58(9)	& $-$18(6) 	& 69(12) 
			& 1.815(8) 	& $-$3.4(6)	& 53(11) 
				& $-$100(2) \\ 
MC without discharge	& 11.3(4) 
		& 42(9) 	& $-$4(7) 		& 54(15) 
			& 1.815(9) 	& $-$5.3(7) 	& 113(14) 
				& $-$107(2) \\ 
MC, no solenoid 	& -- 
		& 31.2(9) 	& $-$2.7(3) 	& 2.6(2) 
			& 0.8946(9) 	& $-$4.45(2) 	& 12.9(2) 
				& 175.6(4) \\ 
\end{tabular} 
\end{ruledtabular} 
\end{table*}

\begin{figure}[t!] 
	\centering
	\includegraphics[width=8.5cm]{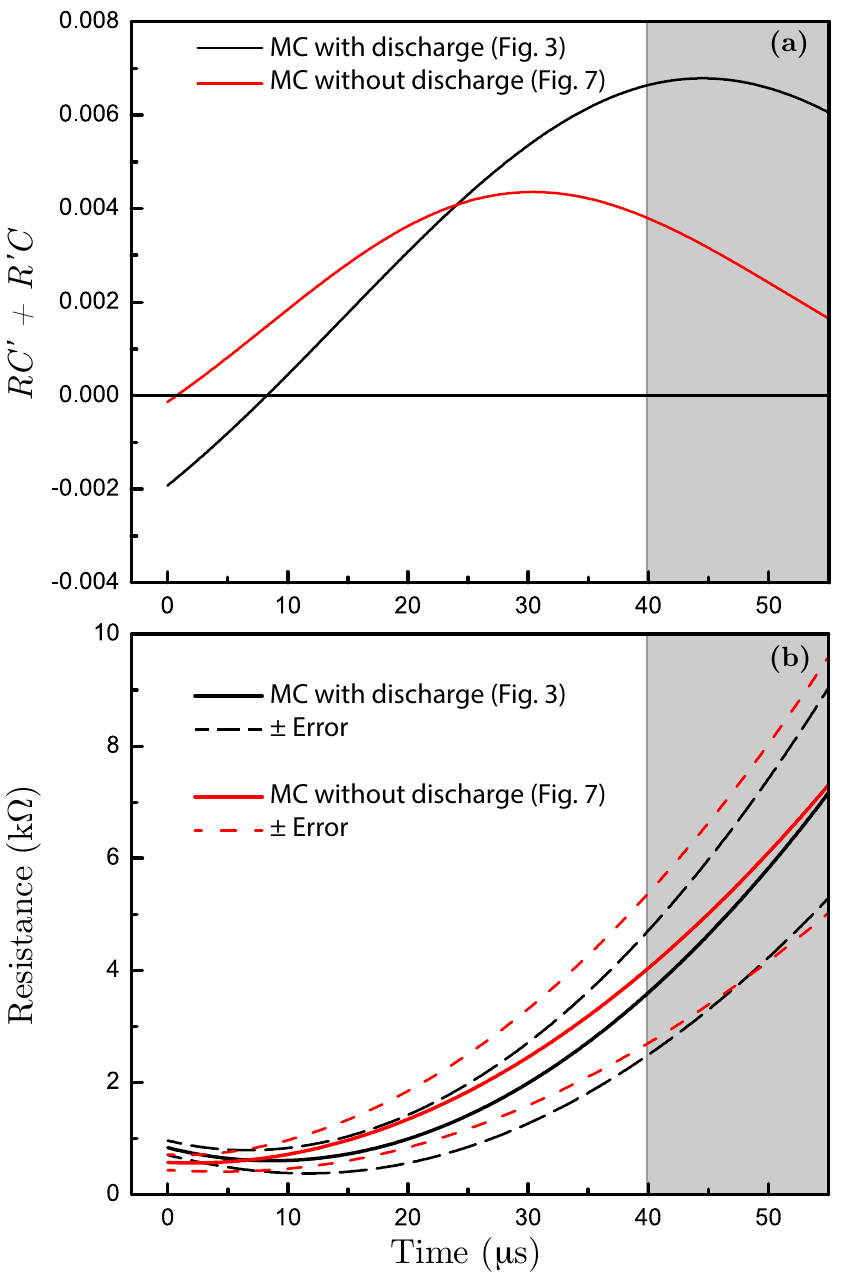}
	\caption{
	Parameter estimation from variable RCL fits of the data in Figs.~3 and 7 using fit parameters in Table \ref{tab:2}. 
	(a) Check of the approximation (\ref{Iapprox}). 
	(b) Estimate of resistance using (\ref{RtI}) that corresponds to the capacitance of (\ref{CtI}) shown in Fig.~9(a). } 
	\label{figSI-Analysis3and11}
\end{figure}

\begin{figure}[t!] 
	\centering
	\includegraphics[width=8.5cm]{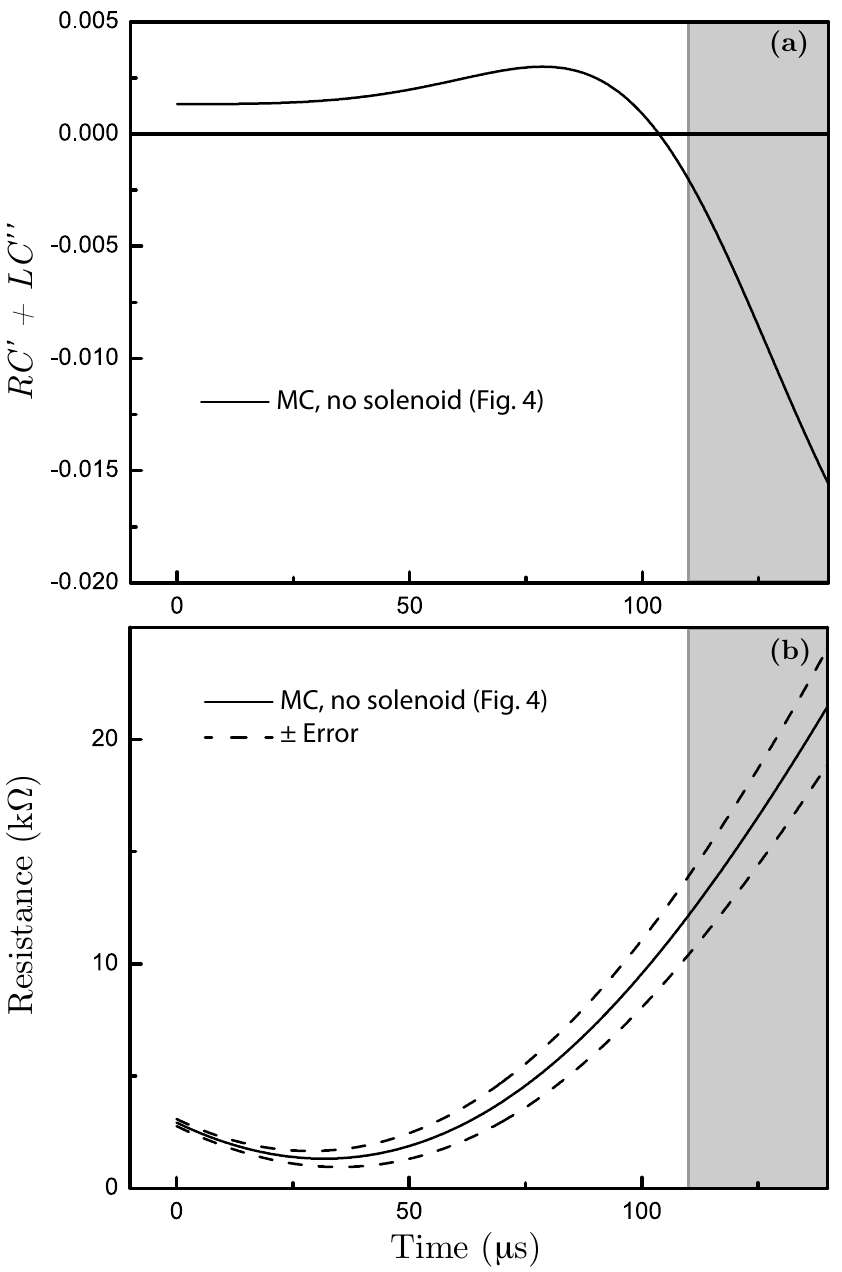}
	\caption{
	Parameter estimation from the variable RCL fit of the data in Figs.~4 using fit parameters in Table \ref{tab:2}. 
	(a) Check of the approximation (\ref{Vapprox}). 
	(b) Estimate of resistance using (\ref{RtV}) that corresponds to the capacitance of (\ref{CtV}) shown in Fig.~9(b). } 
	\label{figSI-Analysis25}
\end{figure}

\subsection{Data analysis} 

Tables \ref{tab:1} and \ref{tab:2} list the fixed and variable RLC fit parameters for the fit curves shown in Figs.~3 and 4 of the paper and the additional fit curves in Fig.~7. 
The additional current data in Fig.~7 corresponds to nearly identical conditions as those in Fig.~3, except that the breakout point was removed to prevent any visible discharge. 
Fig.~8 shows residuals for the fits of Fig.~3, highlighting how the variable RLC fit is an improvement over the fixed RLC fit. 

The fixed RLC fits were of the form  
\begin{align}	\label{FixedRLC}
I(t) = 
\begin{cases}
0, & t < t_0 \\
I_0 + A_0 \, e^{-(t-t_0)/\tau} \sin[\omega (t-t_0)], 
& t \geq t_0. 
\end{cases}
\end{align}
For all current and voltage data, the time axis origin was manually set to match the time of the first signal observed on the oscilloscope. 
Fitting of the current data in Fig.~3(b) and Fig.~7 excluded the first 3 $\mu$s of data because of current transformer bandwidth limitations. 
Likewise, fitting of the voltage data in Fig.~4 excluded the first 2$\mu$s because of capacitive-pickup probe limitations. 

The variable RLC fits used a modified version of (\ref{trialfn}) described above that includes an offset,  
\begin{align}	\label{VariableRLC}
y(t) = 
\begin{cases}
0, & t < t_0 \\
y_0 + A_0 \, e^{-B(t-t_0)} \cos[D(t-t_0) + \phi_0]. 
& t \geq t_0, 
\end{cases}
\end{align}
As before, the time axis origin was manually set to match the time of the first signal observed on the oscilloscope. 
However, here $t_0$ was manually set to zero before fitting.

Fig.~9 shows reconstructed capacitances for the data of Figs.~3, 4, and 7 using the approach described in the previous section. This reconstruction assumes the effective inductance is constant, which is likely a good approximation for the data of Figs.~3 and 7. Unfortunately, no reproducible trend was observed in this and other data that could be attributed to spark discharge, likely because of a larger variability in component effects. 

The solenoid was removed for the data of Fig.~4, increasing the current flowing through the stage inductors and potentially saturating their ferrite cores. Therefore, the reconstruction for the data of Fig.~4 is likely incorrect, and instead represents a fictitious set of time-varying RC parameters with constant inductance that would lead to a similar voltage waveform. 

The capacitance estimation of Fig.~9 used an inductance $L$ of 7.4 mH 
for the current data with the solenoid and of 90 mH for the voltage data without the solenoid. 
Figs.~10 and 11 provide consistency checks of the approximations (\ref{Iapprox}) and (\ref{Vapprox}) required for this estimation, and show corresponding estimated resistances. 
The dashed curves shown in Figs.~9, 10, and 11 were estimated from error propagation of the fit parameter uncertainties.

For Fig.~3, the inferred maximum voltage using the variable RLC fit is 201 kV (using the fixed RLC fit, 228 kV).  
The charging voltage can be approximated as this divided by 18, giving 11.2 kV, which is close to the maximum value of 12.4 kV expected from the power supply in Fig.~2(e) with the Variac set to 140 $V_\text{rms}$ assuming a 120 Hz repetition rate. 
For an erected bank capacitance of 37 pF, this corresponds to a charging energy of at least 0.75 J in the Marx bank. 
Including estimates for the solenoid and terminal contributions given above, the total effective output capacitance is roughly 50.5 pF, which corresponds to a charging energy of 1.0 J.

\subsection{Discharge loading of Tesla transformers}
Unfortunately, the effects of spark discharge loading on Tesla transformers have not been extensively studied.\cite{craven:1997} 
Neither has the related subject of optimizing a Tesla transformer to generate the longest discharge over single or multiple pulses, unlike optimizing to produce the maximum voltage in a single pulse without discharge.\cite{denicolai:2002} These two goals are not identical because nonuniform electric fields normally create the discharge and the longest discharges may form over multiple pulses. 

However, the common approach to maximize the discharge length for a Tesla transformer provides suggestive evidence\cite{EVR:book} for capacitive loading by spark discharge. 
This approach is described in Ref.~\onlinecite{EVR:book} and in online resources for Tesla coil enthusiasts,\footnote{For example: \url{http://www.hvtesla.com/tuning.html}, or more quantitatively, \url{http://www.loneoceans.com/labs/drsstc1/} (Accessed July 2018).} and consists of empirically lowering the uncoupled self-resonant frequency of the primary below that of the secondary until the longest discharge is obtained. 
This approach is often initiated by first attaching a wire to simulate the desired discharge, and then tuning the primary to be resonant with the perturbed secondary before further optimization. 

Presumably, this approach of detuning the primary enables a Tesla transformer to tolerate a larger discharge load during operation.\footnote{This is how online resources for Tesla coil enthusiasts usually justify the approach of tuning the primary self-resonant frequency below that of the secondary to maximize discharge length. These resources typically state that this approach accommodates a dynamically increasing capacitive load from discharge, often referred to as ``streamer loading''} 
However, this has not been extensively studied and it is possible that dynamical effects such as rapid adiabatic passage may contribute.\cite{RAP} 
In contrast, the output voltage of a single pulse is maximized by adjusting the resonant frequencies and coupling strength to particular theoretical values.\cite{denicolai:2002}

\subsection{Further improvements}
The Marx coil apparatus presented here was not optimized for a particular application. 
Improvement is possible using better components, in particular, more stable capacitors and lower-loss switches that allow more control (e.g., solid-state switches or triggered spark gaps). 
Additionally, the MC circuit could be modified to control the output waveform envelope, as is common with Marx generators,\cite{naidu:1995} perhaps to better match the slowly modulated (``beating'') envelope typical of a SGTT. 

Alternatively, another possible opportunity to imitate SGTTs while avoiding their circuit sensitivity may be to drive the solenoid base in series with solid-state switches and feedback, using techniques similar to those in modern Tesla transformer designs,\cite{EVR:book, EVR:minibrute, loneoceans:QCW} instead of driving a solenoid in parallel with a Marx generator as implemented here.


%

\end{document}